# Artificial Intelligence-based Motion Tracking in Cancer Radiotherapy: A Review

Elahheh Salari[1], Jing Wang[1], Jacob Wynne[2], Chih-Wei Chang[1] and Xiaofeng Yang[1*]

[1]Department of Radiation Oncology, Emory University School of Medicine, Atlanta, GA 30322

[2] Department of Radiation Oncology, Icahn School of Medicine at Mount Sinai, New York, NY 10029

**[*]Corresponding author:**

Xiaofeng Yang, PhD

Department of Radiation Oncology

Emory University School of Medicine

1365 Clifton Road NE

Atlanta, GA 30322

E-mail: xiaofeng.yang@emory.edu

## Abstract

Radiotherapy aims to deliver a prescribed dose to the tumor while sparing neighboring organs at risk (OARs). Increasingly complex treatment techniques such as volumetric modulated arc therapy (VMAT), stereotactic radiosurgery (SRS), stereotactic body radiotherapy (SBRT), and proton therapy have been developed to deliver doses more precisely to the target. While such technologies have improved dose delivery, the implementation of intra-fraction motion management to verify tumor position at the time of treatment has become increasingly relevant. Artificial intelligence (AI) has recently demonstrated great potential for real-time tracking of tumors during treatment. However, AI-based motion management faces several challenges including bias in training data, poor transparency, difficult data collection, complex workflows and quality assurance, and limited sample sizes. This review presents the AI algorithms used for chest, abdomen, and pelvic tumor motion management/tracking for radiotherapy and provides a literature summary on the topic. We will also discuss the limitations of these AI-based studies and propose potential improvements.

## 1. Introduction

Radiotherapy aims to deliver a high dose of radiation to treatment targets while minimizing the dose to surrounding healthy tissues. The advent of flattening filter-free (FFF) treatment delivery brought higher dose rate beams and greater normal tissue sparing due to the sharp dose fall-off outside the tumor (1, 2). The FFF technique has widened the therapeutic window, ushering in new radiation delivery techniques such as SRS and SBRT (2, 3). Intrafraction motion monitoring is particularly needed for the SRS and SBRT, where a high dose is delivered to the target in a few fractions, and narrow margins are needed to spare healthy tissues (4). This treatment technique is commonly implemented in the lung, abdomen, and sometimes pelvis, where the efficiency of the treatment can be significantly reduced due to intrafraction respiratory, cardiac, gastrointestinal, and urinary motion during the treatment (5-7). Internal organ movement may cause underdosing or overdosing of targets or normal tissues, potentially causing treatment failure and increasing normal tissue toxicity (8-10). In this setting, real-time tumor tracking techniques are essential to localize targets and ensure accurate treatment delivery without compromising treatment quality due to motion.  Conventional motion management techniques include: using a dynamic multileaf collimator (MLC) to optimize MLC positions based on target motion (11), adjusting the radiation beam and robotic couch according to target movement (12), implanting electromagnetic transponders in the soft tissue to localize tumors or placing transponders on the body surface to monitor the motion (13, 14), utilizing stereoscopic kilovoltage (kV) imaging in conjunction with a six-degrees-of-freedom couch (15), employing an optical surface tracking system using an infrared camera to automatically align patients by tracking infrared (IR) markers on their skin or a rigid template (16), and using ultrasound (US) guidance equipped with a hardware device to hold a US probe in a position that maintains the target within the US imaging field of view during the treatment session (17). Recently, magnetic resonance imaging (MRI) integrated with the linear accelerator has been developed for monitoring intrafractional motion during dose delivery (18). The purpose of this review is not to provide a detailed explanation of each technique, interested readers are directed to Wu et al (19), which contains additional information.

Though various methods for intrafraction motion management have been developed (19), direct detection of the target during the treatment is often not feasible (20). Alternatively, indirect tumor localization facilitated with artificial intelligence (AI) approaches can be used. After decades of development, modern AI approaches can be categorized into machine learning (ML) and deep learning (DL). When applied to motion management, these techniques have successfully analyzed medical images and made motion predictions. AI-based techniques can be applied to several disease sites as well as many imaging modalities including MRI, computed tomography (CT), and US. Due to the wide range of possible applications, numerous motion-tracking strategies have been proposed. Recently, ML approaches integrating radiomics have been developed to analyze medical images. Radiomics is a novel topic in the field of radiology, which extracts mineable quantitative features from medical images. The extracted features contain information on size, shape, and texture from the region of interest and can be used to develop ML models to predict target position (20, 21). Traditional algorithms including artificial neural network (ANN) (22), support vector machine (SVM) (23), light gradient boosting machine (LightGBM) (21), decision trees (DT) (20), random forests (RF) (24) have also been used for predicting tumor position.

In addition to classic ML approaches, many authors have employed DL for real-time tracking of tumors. Convolutional neural network (CNN) is one of the backbone DL architectures used in various medical image/object recognition and classification (25-27). A typical CNN has an initial input layer, and a final output layer with several intervening "hidden" layers connecting the input and output. In CNNs, the hidden layers extract higher-level image "features" from the input image, typically across several resolutions, capturing detail at several spatial scales. A hidden layer may include convolution, pooling, and rectified linear units, in addition to many others. Other more advanced networks such as recurrent neural network (RNN) (28), region convolutional neural network (R-CNN) (29), Siamese networks (30), you-only-look-once (YOLO) (31), long short-term memory (LSTM) (32), and encoder-decoder networks (33) have been introduced for real-time tumor tracking. A detailed taxonomy of DL network architectures is out of the scope of the present article but can be found in a recent review by Wang *et al* (34).

This work aims to review AI-based approaches for tumor tracking in the thoracic, abdominal, and pelvic regions, discuss present limitations, and provide potential solutions for more accurate outcomes.

## 2. Literature Search

This review focuses on intrafraction motion management using classic ML and DL algorithms. To ensure this systematic review is valuable to users, we followed the Preferred Reporting Items for systematic reviews and meta-analyses (PRISMA) (35). For this aim, we only considered peer-reviewed papers as they undergo an evaluation process where journal editors and experts critically assess the article's quality and scientific merit. In this regard, the PubMed search engine was used with a time window from January 2005 to August 2023. The search keywords were limited to “cancer or radiotherapy”, “deep learning or machine learning or artificial intelligence”, “imaging or image-guided radiation therapy or IGRT”, and “motion”. The initial search yielded 173 records. However, after excluding literature reviews and publications not related to medicine, only 87 papers remained. Moreover, a citation search was conducted on other literature resulting in an additional 17 papers; therefore, a total of 104 articles were included in this review study. Figure 1 shows the surging number of yearly peer-reviewed publications containing the terms “artificial intelligence/machine learning/deep learning”, “motion”, “radiotherapy/cancer”, and “images” from 2010 to August 2023 in the PubMed database *(www.pubmed.gov)*. Figures 2 and 3 display the percentage of studies for each treatment site and imaging modality, respectively. In Figure 3. X-ray includes CT, CBCT, 4DCT, MV, and kV fluoroscopic images. Other modalities refer to respiratory gating, electromagnetic transponders, and optical surface monitoring.

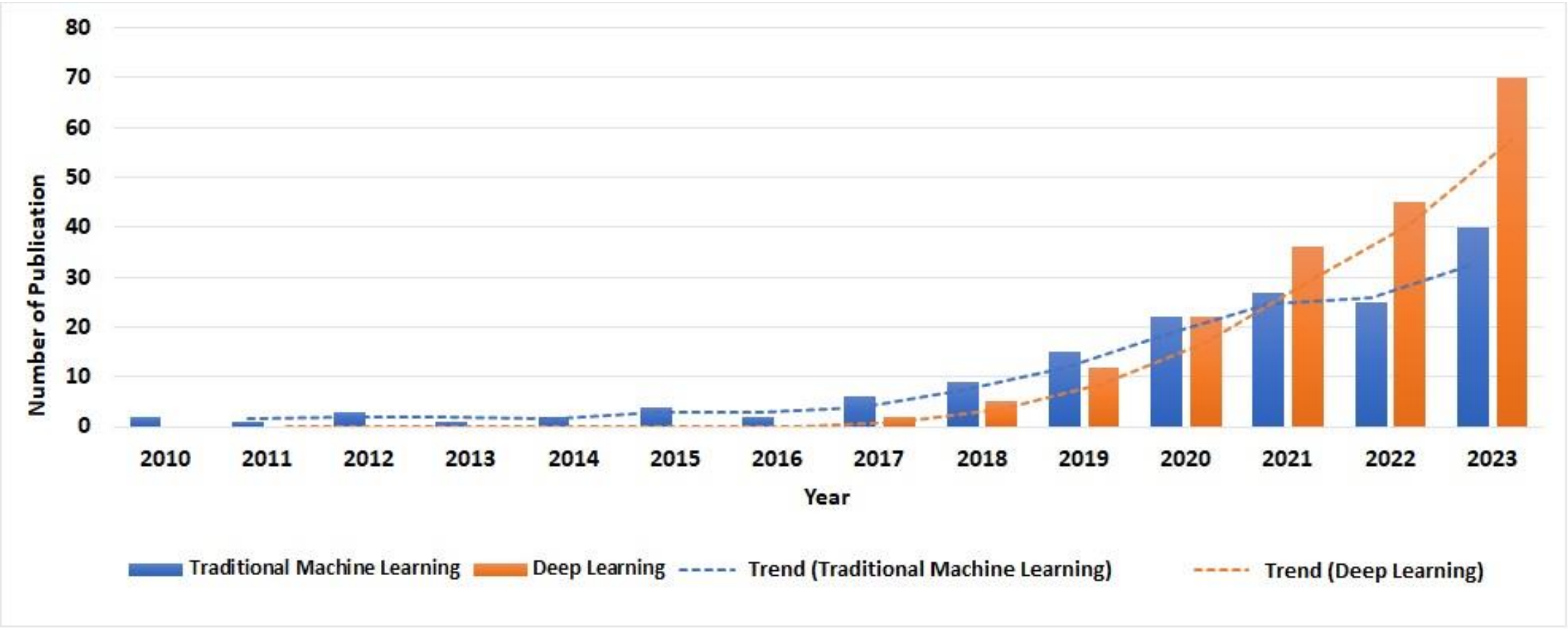

*Figure 1. Number of publications using classic ML & and DL-based motion management since 2010. Deep learning has exponentially increased since 2017. The number of publications in 2023 is an estimation based on the number of publications from January to August 2023.*

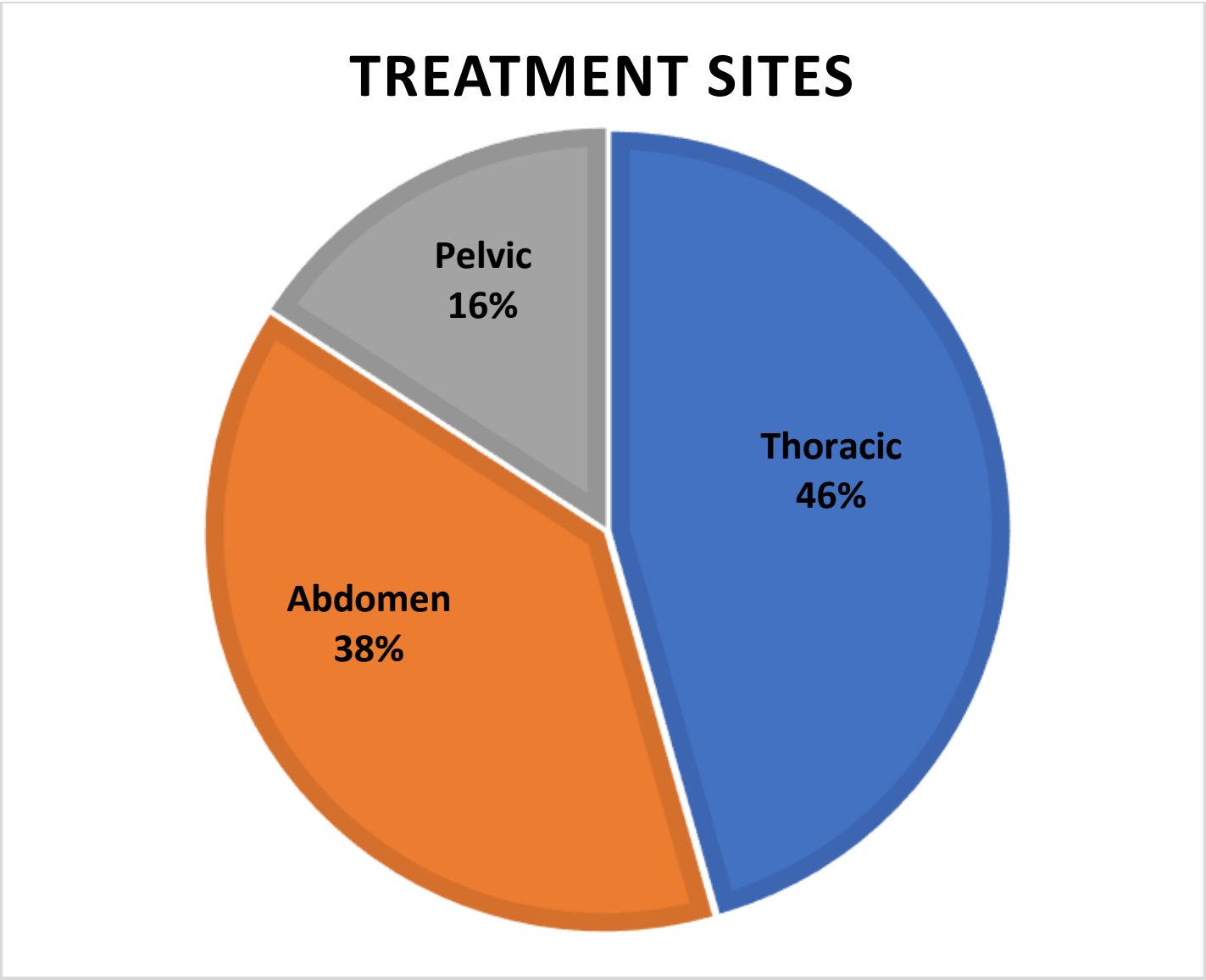

*Figure 2. The percentage of studies conducted on each treatment site using AI-based motion management in PubMed (www.pubmed.gov) since 2005.*

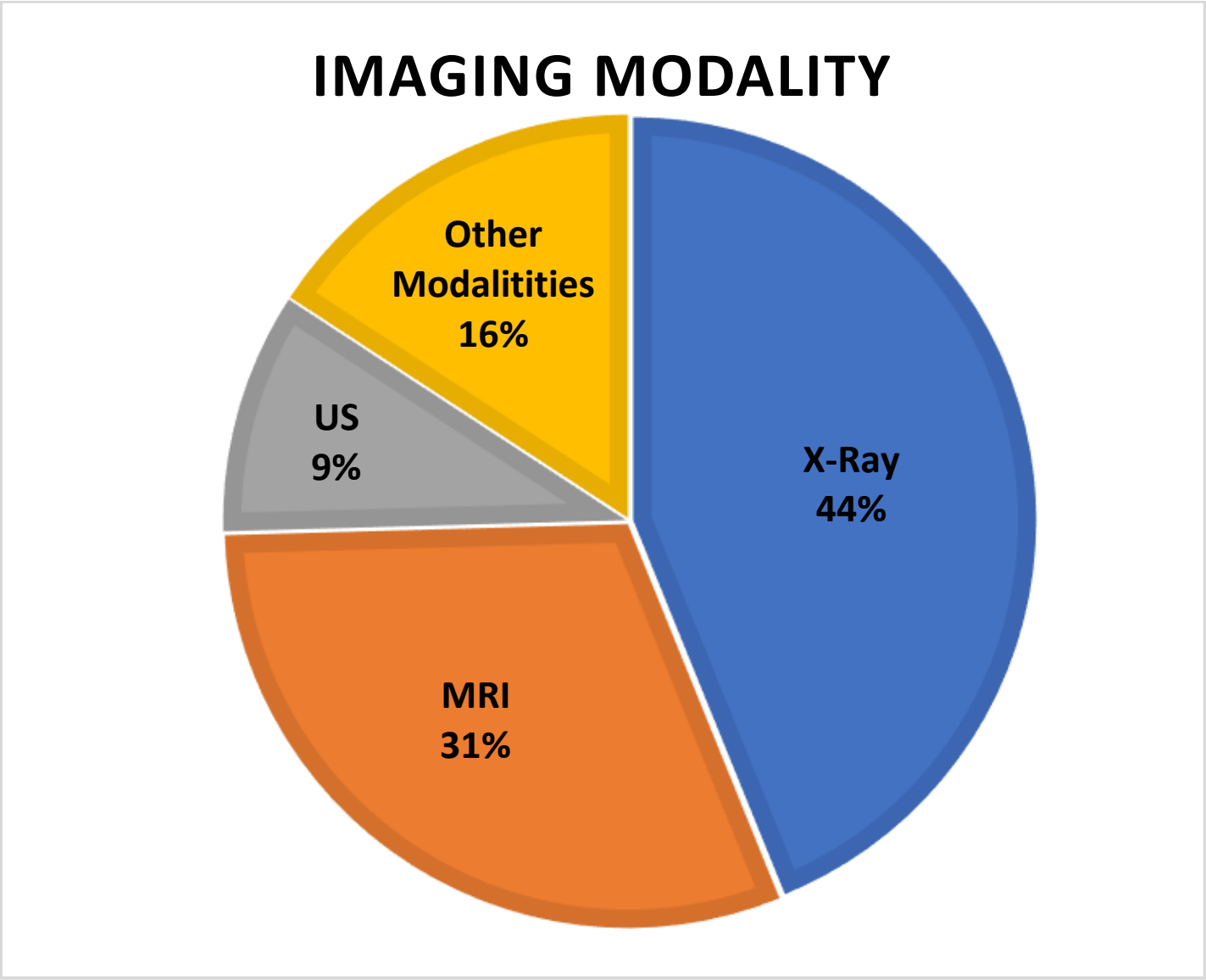

*Figure 3. The percentage of each modality in AI-based motion tracking in PubMed* (www.pubmed.gov) *since 2005. X-ray includes CT, CBCT, 4DCT, MV, and kV fluoroscopic images. Other modalities include respiratory gating, electromagnetic transponders, and optical surface monitoring.*

# 3. Motion management for various treatment sites

## 3.1.Thoracis

According to the American Cancer Society, lung cancer is the leading cause of cancer-related death in the United States (36). Thus, lung cancer has been extensively studied. Radiotherapy is a standard of care in the multi-disciplinary treatment of lung cancer and is being increasingly used but is challenged by intrafraction motion during treatment. Intrafraction motion is primarily a result of respiration, with a lesser contribution of the cardiac cycle to tissue displacement and deformation. Respiration induces organ motion and anatomical shifts which can significantly reduce the accuracy of dose delivery, causing failure of tumor control or normal tissue injury (37). Respiratory gating and breath-hold are common solutions to manage target motion (38), but both techniques provide only a limited representation of the complex respiratory motion pattern (39). To solve this problem, several AI methods across several imaging modalities have been developed. A summary of these approaches in Table 1 demonstrates the majority of authors used X-rays (e.g., CT, CBCT, 4DCT, kV X-ray) in the chest area. When it comes to examining tissue abnormalities in the chest, CT scans are more effective compared to other modalities especially MRI because the chest area is composed of pleural cavities (lungs and pleura) which contain a significant amount of air. This can lower the accuracy and sensitivity of MRI because this modality is a functional imaging technique based on water diffusivity (40). CT scan, on the other hand, is based on X-ray photon absorption which is inversely proportional to the density of the object. Therefore, it can precisely detect lesions in the thoracic regions due to sharp differences in tissue density.

*Table 1. Summary of publication using artificial intelligence for thoracic tumor tracking*

| Author | Year | AI model | Image modality | No. samples | Key findings in results |
|---|---|---|---|---|---|
| Isaksson et al (41) | 2005 | Linear filters, adaptive linear filters, and adaptive neural networks (Feed-forward neural networks) | kV X-ray | 3 (1 pancreatic, 2 lung cancer patients) | The neural network (NN) achieved better tracking accuracy than other algorithms. The NN was able to predict the position of the tumor up to 800 ms in advance. In one case, the linear filter was completely unable to predict the tumor position (nRSME = 100%). |
| Kakar et al (42) | 2005 | Adaptive neuro-fuzzy inference system (ANFIS) | Infrared camera | 11 lung cancer patients | Root-mean-square error (RMSE) for coached patients was 6% and for non-coached patients (breath freely) was 35% over an interval of 20 seconds. |
| Murphy and Dieterich (43) | 2006 | The linear adaptive filter and the adaptive nonlinear neural network | Optical tracking system | 9 lung cancer patients | Predict respiratory signals up to 1 second in advance. The neural network outperformed the linear filter. In some cases, the linear filter was completely unable to adapt to the breathing signal |
| Yan et al (44) | 2006 | ANN | The signal of a simulator and IR camera | 4 patients | The target position can be predicted using external surrogates if the correlation between internal target motion and external marker signals is consistent. |
| Zhang et al (45) | 2007 | Principal component analysis (PCA) | CT | 4 lung cancer patients | The average discrepancies between the predicted model and ground truth were 1.1±0.6 mm LR, 1.8±1.0 mm AP, and 1.6±1.4 mm SI. |
| Cui et al (23) | 2008 | SVM | kV X-ray | 5 lung cancer patients | The SVM can predict the gating signal to deliver the dose to the target with tumor coverage greater than 90%. |
| Lin et al (46) | 2009 | ANN, SVM | kV X-ray | 9 lung cancer patients (ten fluoroscopic image sequences) | ANN approach is more accurate than SVM in terms of classification accuracy and recall rate. The ANN achieved higher accuracy than SVM (96.3±1.6 vs 94.9±1.7). The average running time for ANN is less than SVM (6.7 ms vs. 11 ms). |
| Lin et al (47) | 2009 | Linear regression (LR), two-degree polynomial regression, ANN, and SVM | kV X-ray | 10 lung cancer patients | Two-degree polynomial regression tends to be overfitted. ANN performs better and is more robust than the other models. ANN achieved the lowest localization errors within all models. |
| Riaz et al (48) | 2009 | Multidimensional adaptive filter, SVM | Optical tracking system | 14 lung cancer patients | The SVM prediction results are more accurate than the other model with the RSME equal to 1.26. The RMSE of the multidimensional adaptive filter was 1.71. |
| Torshabi et al (49) | 2010 | ANN and Fuzzy logic | Cyberknife® Synchrony | 15 chest cases and 4 abdominal cases (10 worse cases, 10 control cases) | The best performance was obtained using Fuzzy logic algorithms. The result of ANN was comparable to Cyberknife Synchrony and the calculation time of ANN was higher than other models. The error reduction with respect to Synchrony®, measured at the 95% confidence level is 10.8% for the fuzzy logic approach and 8.7% for ANN. |
| Cervino et al (50) | 2011 | ANN, template matching algorithm with surrogate tracking using the diaphragm | MRI | 5 healthy volunteers | The matching approach detects the target position more accurately than the ANN model. |
| Krauss et al (51) | 2011 | LR, NN, kernel density estimation, and SVM | kV X-ray | 12 breathing data | When considering all sampling rates and latencies, the observed prediction errors normalized to errors of using no prediction for NN, SVR, LR, and KDE were 0.44, 0.46, 0.49, and 0.55 respectively. |
| Li et al (52) | 2011 | PCA | 4DCT | Two phantoms and 11 image sets from eight patients. | The modeling error was within 0.7±0.1 mm. The mean 3D error was 1.8 ± 0.3 mm. |
| Fayad et al (53) | 2012 | PCA | 4DCT and synchronized RPM signal | 10 lung cancer patients | The model is substantially accurate when it includes both phase and amplitude data with the model error of 1.35±0.21 mm. |
| Yun et al (22) | 2012 | ANN (feed-forward neural network) | MRI (MR-linac) | 29 lung cancer patients | For 120–520 ms system delays, mean RMSE values of 0.5–0.9 mm (ranges 0.0–2.8 mm from 29 patients) were observed. |
| Torshabi et al | 2014 | ANFIS | Cyberknife® Synchrony | 10 (lung and pancreas cancer patients) | The ANFIS model was able to decrease tumor tracking errors significantly compared with the ground truth database and even their previous study (49). |
| Li et al (54) | 2015 | MLR | 4DCT | 11 lung cancer patients (2 scans for each patient so 22 4DCT in total) | The ML model was able to predict the diaphragm motion with acceptable error (0.2±1.6 mm). |
| W Bukhari and S-M Hong (55) | 2015 | An extended Kalman filter (LCM-EKF) to predict the respiratory motion and a model-free Gaussian process regression (GPR) to correct the error of the LCM-EKF prediction. | Gating system | 31 patients (304 traces of respiratory motion) | This model reduced the root-mean-square error by 37%, 39%, and 42% for a duty cycle of 80% at lookahead lengths of 192 ms, 384 ms, and 576 ms respectively compared to the non-correction method. |
| Yun et al (56) | 2015 | ANN | MRI (MR-linac) | One phantom and four lung cancer patients | The phantom study yields a mean Dice similarity index (DSI) of 0.95–0.96, and a mean Hausdorff distance (HD) of 2.61–2.82 mm. The mean DSI of 0.87–0.92, with a mean HD of 3.12–4.35 mm for the patient study. |
| Bukovsky et al (57) | 2015 | Multilayer perceptron (MLP), Quadratic Neural Unit (QNU) | 3D time series of lung motion | 356 simulations for QNU and 2475 for MLP | The mean absolute error was 0.987 mm and 1.034-1.041 mm for QNU and MLP respectively. The QNU model provided better results with a mean absolute error of 0.987 and was faster than MLP. |
| Park et al (58) | 2016 | Fuzzy deep learning (FDL), CNN, Hybrid motion estimation based on extended Kalman filter (HEKF) | CyberKnife Synchrony | 130 lung cancer patients | FDL showed fewer variations than CNN and HEKF.<br>The RMSE of FDL was 30 % better than CNN and HEKF.<br>The average computing time using a central processing unit (CPU) for FDL, CNN, and HEKF was 1.54±5.01 ms, 254.32±11.68 ms, and 253.56±10.74 ms respectively. |
| Teo et al (59) | 2018 | A 3-layer perceptron neural network | MV images | 47 (27 in group 1 and 20 in group 2, unseen data) | The average MAE (group 1) = 0.59±0.13 mm, and the average MAE (group 2) = 0.56±0.18 mm.<br>The average RMSE (group 1) = 0.76±0.34 mm and for group 2 was 0.63±0.36 mm. |
| Terunuma et al (60) | 2018 | CNN | 3D CT | For training and testing data sets, 2000 and 300 pairs of model and supervised images were used respectively | The proposed method achieved accurate tumor tracking of low visibility tumors of over 0.95 based on the Jaccard index, and accurate tumor tracking with an error of approximately 1 mm. |
| Edmunds et al (61) | 2019 | Region-CNN (R-CCN) | CBCT | 10 lung cancer patients (3500 raw CBCT projection images) | The performance of the model was lower at lateral angles when larger amounts of fatty tissue obstructed the view of the diaphragm.<br>The model could estimate the diaphragm apex positions with a mean error of 4.4 mm. |
| Jiang et al (62) | 2019 | A non-linear autoregressive model with exogenous input (NARX) | Gating system | 7 lung cancer patients | Mean ± standard deviation was 82.32±17.93%, 80.52±18.00%, and 79.77±18.42% of three different prediction horizons, 600 ms, 800 ms, and 1 s respectively. |
| Lin et al (24) | 2019 | A model was a combination of four base machine learning algorithms such as the RF, MLP, LightGBM, and XGBoost. | 4DCT and the Electronic Health Record | 150 lung cancer patients | The maximum MAE and RMSE were in the superior-inferior (SI) direction with 1.23 mm and 1.70 mm respectively. |
| Hirai et al (63) | 2019 | DNN | 4DCT | 5 lung cancer patients and 5 liver cancer patients. Each 4DCT contains 10 respiratory phases | Averaged track accuracy was 1.64±0.73 mm<br>Accuracy for liver cases was 1.37±0.81mm and for lung cases was 1.9±0.65mm<br>Computation time was less than 40 ms |
| Lei et al (7) | 2020 | The transNet model consists of three modules (encoding, transformation, and decoding modules) | 3D CT generated from 2D CT | 20 lung cancer patients | The mean value of the center of mass distance between manual tumor contours on the ground images and corresponding 3D CT images derived from 2D projection was 1.26 mm, with a maximum deviation of 2.6 mm.<br>The peak signal-to-noise ratio was 15.4 ± 2.5 decibel (dB) and the structural similarity index metric within the tumor region of interest was 0.839 ± 0.090. |
| Mori et al (64) | 2020 | DNN | 4DCT derived from 3DCT | Train set: 2420 thoracic 4DCT from 436 patients<br>Test set: 20 lung cancer patients | The averaged tracking positional errors were 0.56 mm, 0.65 mm, and 0.96 mm in the X, Y, and Z directions, respectively. |
| Sakata et al (65) | 2020 | Extremely randomized trees (ERT) | 4DCT | 8 lung cancer patients | The average tracking positional accuracy was 1.03 ± 0.34 mm (mean ± standard deviation, Euclidean distance) and 1.76 ± 0.71 mm (95th percentile). |
| Dai et al (66) | 2021 | Markov-like network | US | The Cardiac Acquisitions for Multi-structure Ultrasound Segmentation **(CAMUS)** dataset includes 2D US images from 450 patients. The Challenge on Liver Ultrasound Tracking **(CLUST)** dataset consists of 63 2D and 22 3D image sequences from 42 patients and 18 patients | CLUST dataset: A mean tracking error of 0.70 ± 0.38 mm for the 2D point landmark tracking and 1.71 ± 0.84 mm for the 3D point landmark tracking.<br>CAMUS dataset: A mean TE of 0.54 ± 1.24 mm for the landmarks in the left atrium |
| He et al (67) | 2021 | ResNet generative adversarial network | kV X-ray | 20 Patients (1347 2D kV images thoracic and lumbar regions) | The mean error with the decomposed image and the original DRR was 0.13, 0.12, and a maximum of 0.58, and 0.49 in the x- and y-directions (in the imager coordinates), respectively. |
| Momin et al (29) | 2021 | R-CCN, VoxelMorph, U-Net, networks without global and local networks, and networks without attention gate strategy | 4DCT | First experiment (Training set: 20 4DCT, Test set: additional 20 4DCT)<br>The second experiment (training set: 40 4DCT, test set: 9 additional unseen 4DCT). Each 4DCT contains 10 breathing phases | The method was evaluated against several other networks, including VoxelMorph, U-Net, and networks without global or local networks or attention-gate strategies.<br>The Dice similarity coefficients of experiments 1 and 2 were higher than those achieved by VoxelMorph, U-Net, network without global and local networks, and networks without attention gate strategy. Specifically, experiment 1 achieved a coefficient of 0.86 compared to 0.82, 0.75, 0.81, and 0.81 achieved by the aforementioned methods, and experiment 2 achieved a coefficient of 0.90 compared to 0.87, 0.83, 0.89, and 0.89 achieved by the aforementioned methods. |
| Pohl et al (28) | 2021 | RNN, linear predictor (LP), least mean squares (LMS) | 4D CBCT and 4DCT | 4 lung cancer patients (Chest 3D 16-bit image sequences). Each sequence had 10 3D images of the chest at different phases of breathing | RNN was superior to LP and LMS.<br>The maximum prediction error for RNN, LP, and LMS was 1.51 mm, 1.80 mm, and 1.59 mm respectively.<br>RMSE was 0.444 mm, 0.449 mm, and 0.490 mm for RNN, LP, and LMS respectively.<br>The Jitter of RNN, LP, and LMS was 2.59 mm, 2.59 mm, and 2.63 mm respectively. |
| Terpstra et al (26) | 2021 | A multiresolution CNN called TEMPEST | MRI | 27 lung cancer patients (training set:17, validation set:5, test set: 5). Also, the model was evaluated using the publicly available 4DCT dataset. | Compared to the self-navigation signal using 50 spokes per dynamic (366× undersampling), the model was able to provide more accurate motion estimation results.<br>Deformation vector fields were estimated to be within 200 ms, including MRI acquisition.<br>The target registration error of the model on 4DCT without retraining was 1.87 ± 1.65 mm. |
| Liu et al (68) | 2022 | NuTracker model using MLP | 4DCT | 7 lung cancer patients with gold fiducial markers | The proposed model had 26% and 32% improvement over the predominant linear methods with the mean localization error of 0.66 mm and <1 mm at the $95^{th}$ percentile. |
| Lombard et al (69) | 2022 | LSTM, LR | MRI | Training and validation set: 70 patients from group 1. The test set includes 18 patients from Group 1 and 3 patients from Group 2 | LSTM outperformed compared to the LR model. For the 500 ms forecasted interval, a mean RMSE of 1.20 mm and 1.00 mm were obtained for LSTM, while the LR model yielded a mean RMSE of 1.42 mm and 1.22 mm for the group 1 and group 2 testing sets, respectively. |
| Zhang et al (21) | 2022 | Light gradient boosting machine-based recursive feature elimination with radiomics features | 4DCT | 67 lung cancer patients | Mean ± standard deviation was 0.8±0.126, 0.829±0.14, and 0.864±0.086 for thresholds of 0.7, 0.8, and 0.9 respectively. The specificities were 0.771 ± 0.114, 0.936 ± 0.0581, and 0.839 ± 0.101. The area under the curve (AUC) was 0.837, 0.946, and 0.877, respectively. |
| Hindley et al (70) | 2023 | Voxelmap network | CBCT | 2 lung cancer patients (6120 images for training, 680 images for validation, and 680 images for testing). | The mean errors of 3D tumor motion prediction were 0.1 ± 0.5, -0.6 ± 0.8, and 0.0 ± 0.2 mm in the left-right, SI, and anterior-posterior directions respectively. |
| Huttinga et al (71) | 2023 | Gaussian process | MRI (MR-linac) | One phantom, one healthy volunteer, one patient | The root-mean-square distances and mean end-point-distance with the reference tracking method were less than 0.8 mm for all cases. |
| Lombard et al (72) | 2023 | Classical LSTM network (LSTM-shift), Convolutional LSTM (convLSTM), and convLSTM with spatial transformer layers (convLSTM-STL) | MRI | 88 patients (training:52, validation: 18, and test:21) | The LSTM-shift model was found to be significantly better than other models.<br>The maximum RMSE of LSTM-shift was 1.3±0.6 mm.<br>convLSTM and convLSTM-STL yielded the maximum RMSE of 1.9±1.1 and 1.9±1.0 mm respectively. |
| Zhou et al (73) | 2023 | DNN | kV X-ray | 10 lung cancer patients (2250 digitally reconstructed radiographs) | The mean calculation time was 85 ms per image. The median value for the 3D deviation was 2.27 mm overall.<br>There is a 93.6% chance that the 3D deviation is less than 5 mm. |
| Li et al (20) | 2023 | multilayer perceptron (MLP), wide and deep (W&D), categorical boosting (Cat), light gradient boosting machine (Light), extreme gradient boosting (XGB), adaptive boosting (Ada), random forest (RF), decision tree (DT), logistic regression via stochastic gradient descent (SGD), gaussian naive bayes (GNB), support vector classifiers (SVC), linear support vector classifiers (linearSVC), and K-nearest neighbor (KNN) | CT | 108 lung cancer patients | The best result in terms of AUC was obtained by SVC (0.941). Linear SVC provided the best outcome in terms of sensitivity (0.848). The best specificity results were achieved using MLP (0.936). In general, MLP demonstrated the best classification performance and stability among all models. |

### 3.2.Abdomen

Radiation therapy for gastrointestinal cancers often faces two main physical challenges. The first is the proximity of the tumor(s) to many essential OARs such as the duodenum, stomach, small intestine, kidneys, or spinal cord at the level of the abdomen. The second is the mobility typical of both the target and nearby OARs (74). In the abdomen, respiration, peristalsis, and variable organ filling result in variation in target position and organ deformation. The advent of advanced treatment techniques such as volumetric modulated arc therapy (VMAT) and SBRT solved the first challenge by providing highly conformal 3D dose distributions. However, the effective delivery of such a conformal dose to the target requires careful motion management (31, 74). In addition to gating respiratory and breath hold, using an abdominal compression plate is a primary strategy for motion management in the abdomen. AI can be used to augment motion-tracking techniques due to its capability to assess several aspects simultaneously. The summary of these AI-based approaches can be found in Table 2. which indicates MRI is of particular interest because abdominal regions are composed primarily of soft tissue. US can also be used for monitoring anatomical movements in soft tissue and is particularly effective in monitoring hepatic and pancreatic targets due to site accessibility and the absence of osseous obstruction (18).

*Table 2. Summary of publication using artificial intelligence for abdomen tumor tracking*

| **Author** | **Year** | **Algorithm** | **Image Modality** | **No. Patients** | **Key findings in results** |
|---|---|---|---|---|---|
| Gou et al (75) | 2016 | Dictionary learning model | MRI | 3 pancreatic patients and two healthy volunteers (total of 12 imaging volumes) | The dictionary method improved the auto-segmentation, at least 1 of the auto-segmentation method with Dice's index > 0.83 and shift of the center of the organ was less than equal to 2 mm. |
| Stemkens et al (76) | 2016 | PCA | MRI | Phantom and 7 healthy volunteers (Pancreas and kidney) | An average error of 1.45 mm with a temporal resolution < 500 ms was achieved. |
| Ozkan et al (77) | 2017 | Supporters (proposed by Grabner et al (78)) with Leave-one-out cross-validation | US | 24 2D image sequences of Liver | The results for all targets were a mean of 1.04 mm and 2.26 mm 95% percentile tracking error. |
| Dick et al (79) | 2018 | ANN | 4DCT | 4D extended cardiac-torso (XCAT) phantom | The RMSE for the lung-defined tumor motion was 0.67 mm and for the user-defined tumor motion was 0.32 mm. The RMSE of ANN for mismatched data was 1.63 mm, and for the ground-truth data, the RMSE was obtained at 0.88 mm. |
| Dick et al (80) | 2018 | ANN with Leave-one-out-cross-validation | 4DCT | 4D extended cardiac-torso (XCAT) phantom and 8 liver patients | The averaged RMSE was 1.05±1.14 mm and 2.26±2.4 mm for phantom and patients' data respectively. |
| Huang et al (81) | 2019 | k-dimensional-tree-based nearest neighbor search | US | 57 different anatomical features were acquired from 27 sets of 2D ultrasound sequences. | The mean tracking error between manually annotated landmarks and the location extracted from the indexed training frame is 1.80 ± 1.42 mm. Adding a fast template matching can reduce the mean tracking error to 1.14 ± 1.16 mm. |
| Huang et al (82) | 2019 | FCN with convolutional LSTM (CLSTM) | US | The train set was 25 and the test set was 39 liver cancer patients | The mean and maximum tracking error were 0.97±0.52 mm and 1.94 mm respectively.<br>The tracking speed using GPU was from 66 to 101 frames per second. |
| Zhao et al (83) | 2019 | A patient-specific region-based convolutional neural network (PRCNN) | kV X-ray | 2 pancreatic cancer patients (2400 DRR datasets) | The mean absolute difference between the model-predicted and the actual positions < 2.60 mm in all directions.<br>Lin's concordance correlation coefficients between the predicted and actual positions were > 93%. |
| Liang et al (84) | 2020 | FCN | Cyberknife® Synchrony | 13 liver cancer patients (5927 images) | The mean centroid error between the predicted and the ground truth was 0.25±0.47 pixels on the test dataset. The maximum mean translation was seen in the SI direction with 13.1±2.2 mm. |
| Liu et al (85) | 2020 | SVM | Cyberknife® Synchrony | 148 liver cases and 48 cases of other anatomical sites (e.g., Kidney, pancreas) | The sensitivity, precision, specificity, F1 score, and accuracy are 0.81 ± 0.09, 0.85 ± 0.08, 0.80 ± 0.11, 0.83 ± 0.06, and 0.80 ± 0.07, respectively.<br>An AUC of 0.87±0.05 was achieved. |
| Roggen et al (86) | 2020 | Mask R-CNN | CBCT | Train set: 12 abdominal cancer patients (903 images).<br>Test set: 1 patient (49 images) and one phantom with vertebrae | This paper presents a fast DL model for detecting landmarks in vertebrae and evaluates its accuracy in detecting 2D motion using projection images taken during treatment.<br>The proposed network was able to detect the motion in both translational and rotational variations, with sub-millimeter accuracy. |
| Terpstra et al (87) | 2020 | The convolutional neural network called SPyNET | MRI | 7 abdomen, 40 liver, 62 kidney, and 26 pancreas cancer patients (200 images) | Combining non-uniform fast Fourier transform with SPyNET resulted in acceptable performance for 25-fold accelerated data, yielding an imaging frame rate of 25 Hz while keeping the RMSE within 1 mm. |
| Bharadwaj et al (30) | 2021 | Upgraded Siamese neural network using Linear Kalman filter (LKF) | US | CLUS | The proposed model enhanced the Siamese neural network by resolving the constant position model issue and improving robustness.<br>In order to improve the original architecture, LKF was added to include the missing motion model. |
| Romaguera et al (88) | 2021 | Encoder-decoder network | MRI, US | MRI: 25 volunteers, and 11 patients diagnosed with hepatocellular carcinoma. US: 20 volunteers. | The model used image surrogates for volumetric prediction and yielded mean errors of 1.67 ± 1.68 mm and 2.17 ± 0.82 mm for unseen MRI and US patient datasets respectively. |
| Shao et al (89) | 2021 | U-Net | CBCT | 34 liver cancer patients (train set:24, test set:10) | The model obtained the mean center of mass error of 4.7±1.9 mm, 2.9±1.0 mm, and 1.7±0.4 mm,<br>the average DICE coefficients of 0.60±0.12, 0.71±0.07, and 0.78±0.03, and<br>the mean Hausdorff distances of 7.0±2.6 mm, 5.4±1.5 mm, and 4.5±1.3 mm, for<br>2D-3D, 2D-3D deformable registration with biomechanical modeling, and DL model prediction with biomechanical modeling techniques respectively. |
| Wang et al (32) | 2021 | LSTM, SVM | 4D US and template matching to track the motion | 7 volunteers | LSTM was superior to SVM with RMSE less than 0.5 mm at a latency of 450 ms for the prediction of respiratory motion and internal liver motion of < 0.6mm. |
| He et al (90) | 2021 | ResNet generative adversarial network (ResNetGAN) | CT | Train set: 20 patients (1347 2D kV thoracic and lumbar region).<br>Test set: 4 patients (226 2D kV images) | The model was able to enhance image quality for submillimeter accuracy.<br>The decomposed spine image was matched with the ground truth with an average error of 0.13, 0.12, and a maximum of 0.58, and 0.49 in the x- and y-directions respectively. |
| Liu et al (91) | 2022 | DNN | MRI | 7 liver cancer patients | The proposed model increased the image quality in terms of structural similarity index, peak signal noise ratio, and mean square error.<br>The median distance between the predicted model and the ground truth in the SI direction was 0.4±0.3 mm and 0.5±0.4 mm for cine and radial acquisitions respectively. |
| Shao et al (92) | 2022 | RegNet, KS-RegNet-nup, and KS-RegNet | MRI | Eight cases from an open-access multi-coil k-space dataset (OCMR) were used for the cardiac dataset,<br>9 liver cancer patients for the abdominal dataset were selected. | KS-RegNet was found to be better, and more stable performance compared to other models. |
| Shao et al (93) | 2022 | Graph neural network or GNN | 4DCT | 10 liver cancer patients and each patient had 10 respiratory phases | The mean localization error was less than 1.2 mm.<br>This indicates the potential of the model for tumor tracking. |
| Ahmed et al (31) | 2023 | CNN, YOLO, and CNN-YOLO | kV X-ray | training data (44 fractions, 2017 frames). Test data (42 fractions, 2517 frames) | The MAE and RMSE of all 3 models were less than 0.88±0.11 mm and 1.09±0.12 mm respectively. |
| Dai et al (94) | 2023 | CNN | Optical surface monitoring, and kV X-ray | 7 liver cancer patients | The maximum MAE and RMSE were observed in the SI direction (3.12±0.80 mm, and 3.82±0.98 mm respectively). |
| Hunt et al (95) | 2023 | DL (VoxelMorph and U-Net), Affine, b-Spline, and demons | MRI | 21 patients with abdominal or thoracic tumors (>629000 frames from 86 treatment fractions) | The DL model provided better results compared to conventional methods.<br>The RMSE was 0.067, 0.040, 0.036, and 0.032 for affine, b-spline, demons, and DL respectively. |
| Shao et al (96) | 2023 | DL-based framework (Surf-X-Bio) | kV X-ray and surface imaging | 34 liver cancer patients | The Surf-X-Bio can precisely monitor liver tumors through a combination of surface and x-ray imaging compared to surface-image-only and x-ray only models. |

## 3.3.Pelvis

According to the American Cancer Society, prostate cancer is the most common malignancy of men in the US, accounting for 27% of the total diagnoses of all sites (36). Thus, prostate cancer has been widely studied. Radiotherapy is indicated as a primary and salvage treatment for prostate cancer in both early and advanced diseases. The major challenge of prostate radiation therapy is unpredictable intrafraction prostate motion due to variable rectal and bladder filling. Target position uncertainties in prostate cancer radiotherapy are usually addressed by assigning a margin around the target volume (97-99). However, without continuous monitoring and intervention, intrafraction motion can cause a geographic miss in approximately 10% of SBRT prostate therapy cases (5). Various strategies have been proposed for intrafraction motion monitoring such as US, kV/kV imaging, infrared cameras, implanted fiducial markers with in-room imaging, CBCT, and MRI (98, 99). Among these, fiducial markers used in conjunction with X-ray imaging is most commonly employed, allowing for real-time target localization and tracking (100, 101). However, even with appropriate use, marker migration within and outside of the prostate can reduce the dosimetric coverage of the target volume and increase the dose to OARs (101). AI-based models can be a potential tool to resolve complications associated with marker-based techniques. Table 3 presents a summary of AI approaches to motion management in pelvic targets.

*Table 3. Summary of publication using artificial intelligence for pelvic tumor tracking*

| **Author** | **Year** | **Algorithm** | **Image Modality** | **No. Patients** | **Key findings in results** |
|---|---|---|---|---|---|
| Zhao et al (102) | 2019 | DNN | CT | 10 prostate cancer patients | Differences between the positions, predicted by DNN and ground truth positions are (mean ± standard deviation) 1.58±0.43 mm, 1.64±0.43 mm, and 1.67±0.36 mm in anterior-posterior, lateral, and oblique directions, respectively. |
| Zhu et al (103) | 2019 | Deep convolutional neural network | US | 83 image pairs from 5 prostate cancer patients | CNN registration errors were < 5 mm in 81% of the cases. While manual registration errors were less than 5 mm in 61% of the cases. Also, advanced normalized correlation registration errors were < 5 mm only in 25% of the cases. |
| Mylonas et al (104) | 2019 | CNN(AlexNet) | kV X-ray | Training set (a phantom and 5 intrafraction images from three prostate cancer patients). Test set (12 fluoroscopic intrafraction images from 10 prostate cancer patients). | The method was effective for continuous fluoroscopic imaging where the markers were in the tracking window in subsequent image acquisition. |
| Amarsee et al (105) | 2021 | You Only Look Once (YOLO) convolutional neural network | kV X-ray | One phantom | The fiducial marker seeds were successfully detected in 98% of images from all gantry angles; the variation in the position of the seed center was within ± 1 mm. The percentage difference between the ground truth and the detected seeds was within 3%. |
| Fransson et al (97) | 2021 | Principal component analysis (PCA) (unsupervised machine learning) | MRI (MR-Linac) | 9 healthy male volunteers | The cumulative variance of the eigenvalues from the PCA showed that 50% or more of the motion is explained in the first component for all subjects. |
| Nguyen et al (106) | 2021 | Kalman Filter framework | kV X-ray | 17 prostate cancer patients (536 trajectories) | The maximum RMSE (without noise) was obtained at 0.4±0.1 mm. With noise, the RMSE was 1.1±0.1 mm. |
| Motley et al (101) | 2022 | You Only Look Once (YOLO) | kV X-ray | 14 prostate cancer patients (~ 20,000 pelvis kV projection images) | The detection efficiency of the model was 96% with an RMSE of 0.3 pixels. |
| Chrystall et al (100) | 2023 | CNN | MV-based IGRT | 29 prostate cancer patients | The AUC of 0.99, the sensitivity of 98.31%, and the specificity of 99.87% were achieved. The mean absolute geometric tracking error was 0.30 ± 0.27 for lateral and 0.35 ± 0.31 mm for SI directions of the MV images. |

Tables 1-3 are not exhaustive; rather, representative publications are provided for illustration within each table. Studies covering several treatment sites are listed only once.

## 4. Artificial Intelligence for motion tracking in radiotherapy

### 4.1. Machine learning-based motion tracking

ML is a subfield of AI that utilizes algorithms to analyze input data and learn from them to provide recommendations and decisions. The typical workflow of classic ML includes data collection, data pre-processing, dataset (training, validation, and test) building, evaluation, and finally, deployment to production. In radiotherapy, classic ML methods can assist in analyzing different aspects of target motion to predict future positions and optimize treatment delivery. Several classic ML models have been proposed for tumor motion tracking. An ANN approach has been used in several studies to predict target position during treatment delivery. Isaksson et al (41) use the positions of external surrogate markers as an input for adaptive neural networks. The study demonstrated that the adaptive neural network provides a more accurate estimation of tumor position compared to fixed and adaptive linear filters. A subsequent study conducted by Murphy and Dieterich (43) confirmed the advantage of an adaptive neural network by comparing it against the linear adaptive filter. Krauss *et al* (51) performed a study on 12 samples of breathing data using linear regression (LR), kernel density estimation, SVM, and ANN. They indicated there were small differences between the models (Table 1). To improve the accuracy of ANN, some authors used an adaptive neuro-fuzzy inference system (ANFIS) which combines the benefit of both neural network and fuzzy logic systems (42, 49, 58). Kakar *et al*(42) used ANFIS to predict respiratory motion both more precisely and more quickly. Yan *et al*(44) developed a technique using ANN to predict tumor position by correlating the internal target position and an external surrogate. The proposed technique assumes a consistent correlation between internal and external movements, allowing for their prediction errors to be correlated with a linear model. In 2008, Cui *et al*(23) proposed that an SVM is a potentially accurate and efficient algorithm for predicting target position. Riaz *et al* (48) analyzed the performance of multi-dimensional adaptive filters versus a support vector machine (SVM) to predict lung tumor motion. They showed performance superior to the SVM model with RMSE < 2 mm at 1-second latency. Lin *et al* (47) further demonstrated the superiority of an ANN approach over SVM (Table 1). An extremely randomized tree (ERT) algorithm can also be used to predict tumor motion. Sakata *et al* (65) trained an ERT for position prediction of lung tumors using digitally reconstructed radiography (DRR) as inputs. They also used sliding window classification to provide a tumor likelihood map. The model was tested on 4DCT of eight patients and yielded an accuracy of $1.0 \pm 0.3$ mm (Table 1). In 2015, Bukovsky *et* *al* (57) presented a combination of quadratic neural unit (QNU) with modification of the Levenberg-Marquardt (L-M) algorithm to improve the accuracy of prediction. The authors achieved a prediction error of less than 1 mm on average of internal tumor position in total treatment time. Moreover, they indicated that QNU with the Levenberg-Marquardt (L-M) algorithm is faster and can yield more accurate results than multilayer perceptron (MLP). Whereas a study conducted by Li *et al*(20) showed that the MLP model had better classification performance and stability for both lung and liver tumors compared to other models. The authors compared thirteen algorithms such as MLP, wide and deep (W&D), categorical boosting, light gradient boosting machine, extreme gradient boosting, adaptive boosting, random forest, decision tree, logistic regression via stochastic gradient descent, Gaussian Naive Bayes, SVM, linear support vector classifiers, and K-nearest neighbor. All models were developed based on radiomic features extracted from CT images of 108 patients with lung cancer and 71 patients with liver cancer. Stemkens *et al* (76) proposed a patient-specific method using a 3D motion model and fast 2D cine-MR imaging to estimate abdominal motion. The motion model was obtained by performing a principal component analysis (PCA) on inter-volume displacement vector fields (DVFs) that were derived from a pre-treatment 4D MRI scan.

As we can see from Tables 1 to 3, the most popular classic ML algorithms for tumor tracking are ANN and SVM algorithms. ANN is of further interest due to its ability to capture both dynamic and structural phenomena, noise suppression, edge detection, and image enhancements (43, 57, 107). These features make it perhaps the best choice among other classic ML techniques for finding useful solutions that require less human intervention. The SVM has been used widely due to its high performance, flexibility, and efficiency for small datasets (108). SVM is useful to deal with nonlinear classification based on a linear discriminant function in a high-dimensional (kernel) space (109). This feature improves its performance in processing nonlinear problems and real-time dynamic predictions (23, 32, 109).

The descriptions and drawbacks of common classic ML approaches are summarized in Table 4.

*Table 4. A list of more common classic ML algorithms used for tumor tracking position.*

| Algorithm | Description | General Drawbacks |
|---|---|---|
| Artificial neural network (ANN) | A typical neural network has three layers including an input layer, a hidden layer, and an output layer. This system can process information and adjust to changing situations to optimize its performance in real time. It yields great accuracy for nonlinear and irregular patterns. It can be used for regression and classification.<br>ANN includes MLP, QNU, Adaptive neuro-fuzzy inference system (ANFIS), Non-linear autoregressive with exogenous, wide, and deep (W&D) | Tend to overfit on small size datasets<br>Memorize the training data and do not generalize well the learned knowledge to new or different data (110, 111) |
| Support vector machine (SVM) | The SVM is a supervised ML algorithm that performs classification and regression by finding the best line or decision boundaries to separate data into classes. The algorithm is a kernel-based model to solve linear and non-linear problems (109). | Strongly depends on the kernel<br>Sensitive to noise<br>Not suitable for large dataset |
| Decision Tree (DT) | The DT is a supervised ML algorithm that builds trees during training time over the entire dataset (112) | Sensitive to change<br>It requires a relatively longer time to train the model (112) |
| Gaussian process regression (GPR) | The GPR is a probabilistic supervised ML algorithm to make predictions with uncertainty (113). | Their efficiency may decrease in high-dimensional spaces.<br>The prediction is based on the entire sample/feature information (113) |

## 4.2.Deep learning-based motion tracking

DL, a class of ML, uses a stack of processing layers with non-linear units that extract higher-level features from inputs. The advantage of DL over classic ML is that DL uses artificial neural networks to automatically learn from data and improve performance over time. The multilayered structure of DL enables it to self-train based on inputs and desired outputs (114). The workflow of DL is similar to classic ML and starts with data acquisition and preprocessing followed by building and training the model, optimization, evaluation, and predictions or inference. Recent advancements have increased interest in DL-based models for tracking target motion (Figure 1) and several algorithms have been trained for this task across most image modalities (Figure 3). In 2019, Zhao *et al* (102) conducted a study on 10 patients with prostate cancer who underwent either CBCT or orthogonal kV projections. The DL model used two networks: a region proposal network (RPN) and CNN. RPN was used to generate proposals for the region-based CNN to reduce computation time and enable real-time target detection. The two networks share all convolutional features. The study (Table 3) showed that highly accurate tumor localization can be achieved using CNN. However, Park *et al* (58) demonstrated that ANFIS was able to achieve an RMSE of 0.5 ± 0.8 mm using a 192.3 ms prediction, a 30.0% improvement over CNN. According to Park et al, the fuzzy logic component enhances the reasoning ability of the model when dealing with uncertainty (58). A study by Liang et al. (84) used an automated framework to evaluate intrafraction motion in CyberKnife X-ray images. The framework utilizes a fully convolutional network (FCN)-based module to detect fiducial markers and perform semantic segmentation using a U-Net architecture in full-size X-ray images. The 3D positions of the markers can then be reconstructed to evaluate intrafraction motion using a rigid transformation (Table 2). Wang *et al* (32) compared an LSTM approach against an SVM to estimate external respiratory motion and internal liver motion. The LSTM network was found to perform better on all planes. Edmunds *et al* (61) proposed to automatically segment the diaphragm using CBCT images for real-time tracking of lung tumors. A Mask R-CNN (Region Convolutional Neural Network) was trained on 3499 raw CBCT images from 10 patients with lung cancer. No manual intervention was required, and the model was able to track diaphragm motion in real-time with a mean error of 4.4 mm. Several attempts were made to learn a joint mapping between partial views and 3D shapes. These efforts have established a path for connecting partial observations with high-dimensional data within a comprehensively trained deep framework. In 2020, Lei *et al* (7) proposed a novel method named TransNet including encoding, transformation, and decoding modules to draw 3D CT images from 2D projections. Later a fast volumetric imaging method with an orthogonal 2D kV/MV image pair using their modified deep learning method was developed for tumor localization in lung cancer radiation therapy (115). The result of this novel method (Table 1) suggests the feasibility and efficacy of the proposed technique to convert 2D to 3D images which can be a potential solution for real-time lung tumor tracking in SBRT cases. The CT images were generated for one case by three different supervision strategies: 1) supervised by mean absolute error (MAE) and gradient difference (GD) losses (TransNet-v1); 2) supervised by MAE, GD, and perceptual losses (TransNet-v2); 3) supervised by a combination of MAE, GD, perceptual and adversarial losses (TransNet-v3). Figure 4 shows the volumetric image generation based on different X-ray projections acquired at different angles using TransNet-v3.

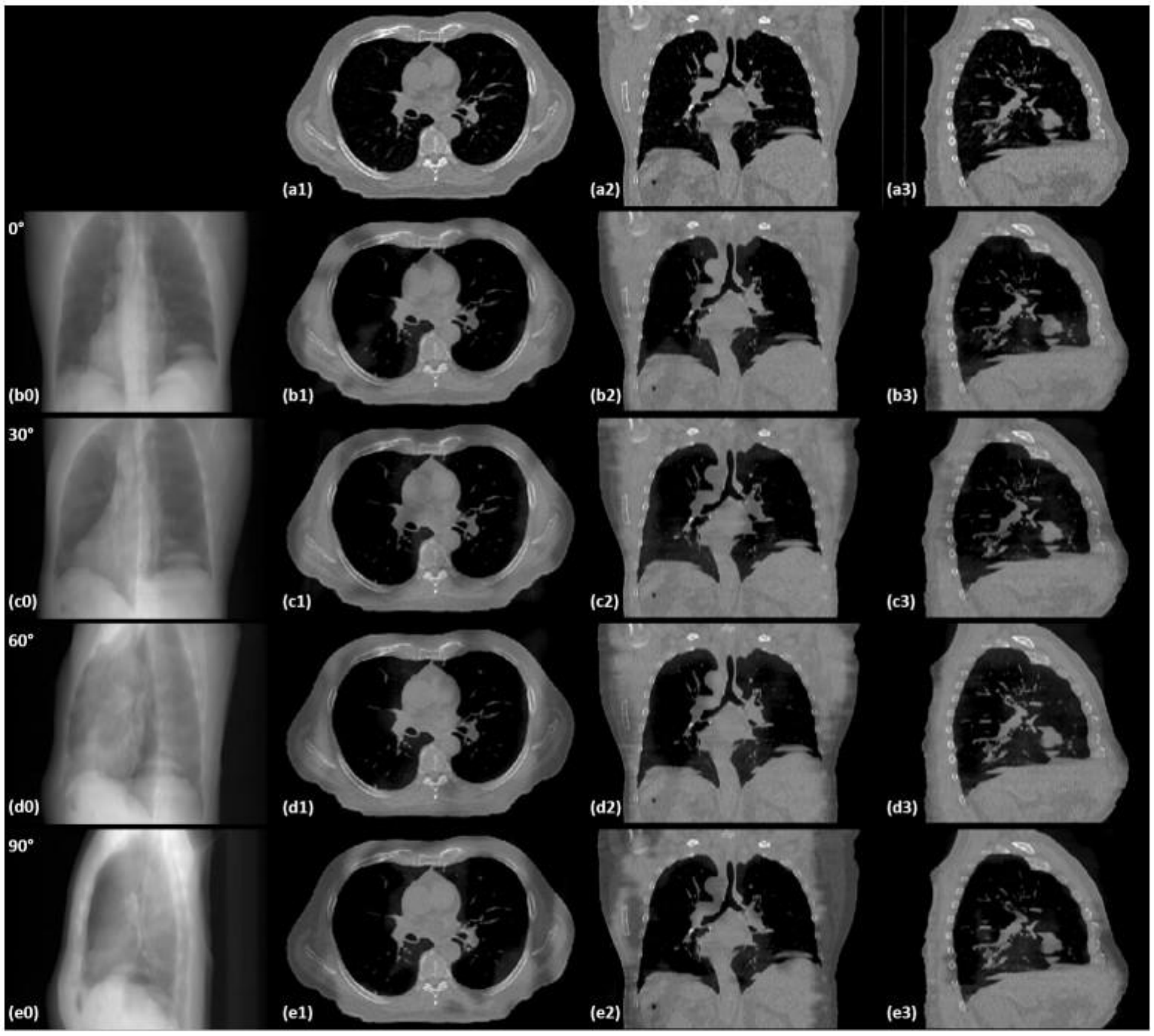

*Figure 4. The 3D CT images are generated from a single 2D kV projection. (a1-a3) are ground truth 3D CT in axial, coronal, and sagittal views. Rows (b-d) demonstrate the projection data and corresponding generated 3D CT images for projection angles 0, 30, 60, and 90, respectively (7).*

Liu *et al* (68) developed a new template network named NuTracker using an MLP comprising eight fully connected layers. The proposed model decomposes 4DCT images into template images and deformation fields using two coordinate-based neural networks to generate predictions from spatial coordinates and surrogate states. Hirai *et al* (63) trained a deep neural network (DNN) to generate a target probability map (TPM) to predict the position of lung and liver tumors. Crops of the target and surrounding anatomy were produced from DRR images. These crops were used to produce the TPM. Accuracy was quantified using the Euclidian distance in 3D space between the calculated and reference tumor position (Table 1). US can provide real-time volumetric images to track intra-fraction motion during radiotherapy. Dai *et al* (66) implemented a generative adversarial Markov-like network (GAN-based Markov-like net) to estimate deformation vector fields (DVFs) from sequential US frames. The positions of the landmarks in the untracked frames were determined by shifting landmarks in the tracked frame based on estimated DVFs (Figure 5). Zhang *et al.* (116) developed a cascade deep learning model with an attention network, a mask region-based convolutional neural network (mask R-CNN), and a long short-term memory (LSTM) network to track real-time liver motion in US image-guided radiation therapy.

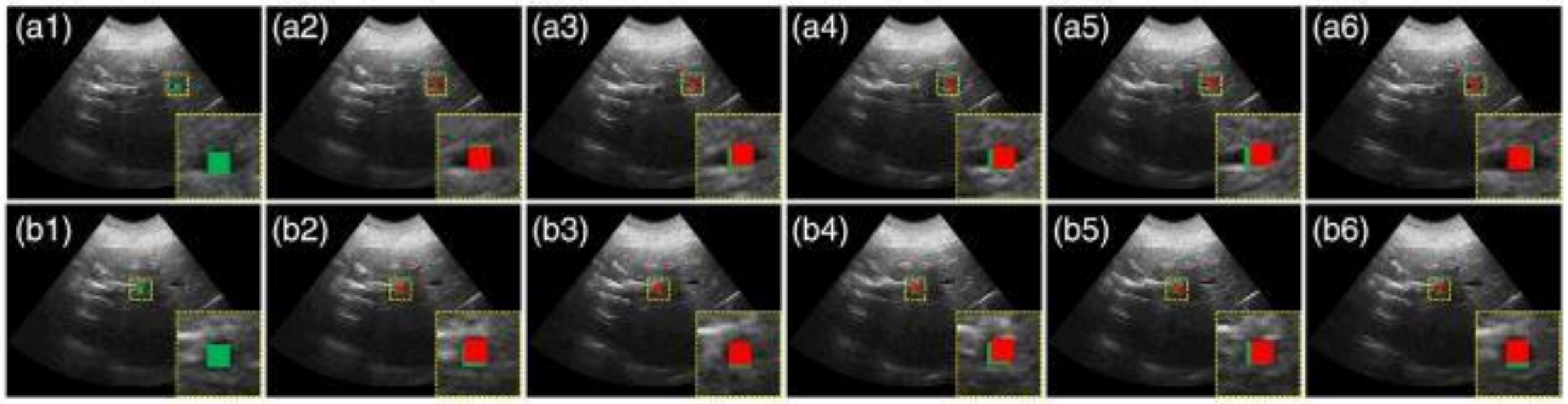

*Figure 5. An illustrative example of 2D point-landmark tracking. a1 and b1 are ground truth and the location of ground truth landmarks are shown in green boxes. (a2-a5) and (b2-b5) are predicted locations and red boxes indicate the predicted landmark positions in the two different landmark tracking (66). (With permission)*

## 5. Discussions

In the past, covering the target with large margins was required to compensate for intrafraction motion during radiotherapy, despite the risk of high-dose delivery to OARs. Techniques like tumor tracking or respiratory gating are now commonly used to maintain target coverage, but real-time monitoring is required to trigger beam on/off signals during gating (18). Intrafraction motion management may use X-ray, MRI, US, and other means to improve the effectiveness of radiotherapy. X-ray is the most common modality compared to other imaging techniques, primarily due to accessibility (Figure 3). X-ray use began with the use of megavoltage beam (MV) portal imaging. However, due to the poor contrast resolution of MV images, in-room kV imaging systems were introduced. In kV images, soft-tissue contrast is higher, particularly when CBCT imaging is used. Planar kV (2D) and CBCT (3D) imaging are now typically available on modern standard-equipped linear accelerators. Moreover, motion-tracking systems equipped with IR-based monitoring can be used in conjunction with X-rays to further improve internal-external correlation (18). For example, the CyberKnife Synchrony system utilizes stereoscopic X-ray imaging to detect implanted clips and correlate them with external surrogates (49). Varian TrueBeam linear accelerators also use respiratory gating and on-board kV imaging to verify internal target anatomy at the start of the gated treatment window, guided by the RPM signal (18). Despite the many advantages of X-ray based imaging techniques, limitations remain. X-ray-based kV and specifically MV imaging reliably capture osseous anatomy but fail in effectively resolving tumor motion within soft tissues (117). CBCT may be used to acquire a volumetric image of the patient's anatomy before treatment but cannot be used during treatment delivery. Regardless, CBCT is prone to severe artifacts, including streaking, beam hardening, and aliasing as well as those due to motion, which severely degrade image quality (118). To minimize motion artifacts during CBCT image acquisition, inspiratory breath-hold was once commonly employed (119); however, modern chest radiotherapy now uses respiratory time-series imaging (4DCT) to capture physiologic motion throughout the breathing cycle (38). Although 4DCT represents a meaningful technological advance in the delivery of radiation therapy, it cannot provide information on variations between breathing cycles or variations occurring on a time scale beyond a magnitude of seconds (120). Using 4DCT/CBCT with implanted fiducial markers for real-time tracking of tumor positions in chest, abdomen, and pelvis regions introduces additional risk associated with possible procedure site infection, fiducial marker migration, non-trivial additional imaging dose exposure, increasing patient expense, and discomfort (121, 122). Moreover, if fiducial markers are obscured by high-density material like bone, surgical clips, or stents, their value is significantly diminished (31).

The second imaging modality for real-time tumor tracking in clinical practice is MRI due to the integration of MR imaging with commercially available linear accelerators. The introduction of the MRI-linear accelerator (MR-linac) enables imaging both before and during treatment with greater soft tissue contrast relative to X-rays, making it easier to differentiate targets from normal tissue (97). Notably, MR-linac does not require the implantation of fiducial markers or the delivery of extra doses to the patient. However, its use is limited by the size of the imaging bore and is contraindicated in the presence of ferrous metal implants or cardiac pacemakers/defibrillators.

US imaging is another well-established imaging technique that can provide real-time volumetric images to track intra-fraction motion without ionizing radiation (81). Yet, due to the poor penetration of US waves into deeper tissues, US cannot be used reliably in clinically important regions such as the skull and thorax (123).

To resolve the aforementioned issues, the use of AI to track the target motion during treatment on the chest and abdomen treatment sites, where SRS/SBRT are commonly used has been extensively studied (28, 29, 56, 94, 101). Mainly because, AI-based techniques have shown great potential to address this question of how information—or lack thereof—collected during imaging can be used in real-time motion tracking (Tables 1-3).

### 5.1.Transition to Deep Learning

As illustrated in Figure 1, there has been a shift in AI-based motion-tracking research from ML to DL over the past decade. The drivers of this transition include the presence of large, high-quality, publicly available labeled datasets, along with the rapid advances in parallel graphic processing unit (GPU) computing, enabling more time-efficient computing and image analysis (61, 124). Before the emergence of powerful computers, CNNs, as a component of larger DL networks, required a significant amount of time to make predictions and provide results when using a central processing unit (CPU) (Table 1) (58). Classic ML techniques require greater human input to achieve reasonable results, intrinsically introducing human error and bias that may influence a study's outcome (114). For instance, in classic ML with radiomics workflow, the features are extracted manually from images and then models are built based on those features to categorize the object in the image. With DL, the extraction of the relevant features from images and modeling steps are automatic (114, 124). Compared with classic ML approaches, DL-based methods are more generalizable, as the same network and architecture used for one image modality can be applied to different pairs of image modalities with minimal adjustments (26, 34). This enables fast translation to multiple clinically useful imaging modalities. Unlike classic ML methods that tend to reach a plateau at a certain level of performance when more examples and training data are added to the network (125), DL networks often continue to improve as the size of data increases (114). DL-based methods have gained significant research and clinical interest in medical imaging generally and radiotherapy particularly due to these specific advantages.

## 5.2.Current Limitations and Potential Solutions

Despite the proliferation of AI-based solutions in healthcare, efforts toward standardization and regulation are lacking (126, 127). Many AI techniques fail to meet expectations for clinical utility, requiring significant time or computing resources to complete complex calculations. Additionally, the presence of bias in human-led data collection and model training, coupled with a lack of standardized reporting for research and clinical validation results makes it difficult to reproducibly apply these techniques in practice (128).

Furthermore, trained algorithms often fail when applied to different datasets due to limited generalization (resulting from small and homogeneous training data sets) (125, 129). Tables 2-4 indicate that some early studies had a limited number of patients, resulting in inadequate patient datasets for systematic analysis of intrafraction motion. Small sample sizes can produce false-negative results (130) and may yield falsely higher accuracy due to overfitting or random effects (125, 131). Ultimately, the limited size of the training and test sets introduces bias and increases variance in model performance (108, 114). To resolve this issue, many follow Cohen's equations (132) to determine the effect size by calculating the mean and variance. However, there are two methods to calculate the variance: a) population variance, and b) sample variance. To calculate population variance, all data is needed whereas for computing sample variance we only need a portion of it. This difference in variance calculation can affect the outcome of these measurements. Rajput *et al* (125) proposed a new method to evaluate sample size using effect sizes (average and grand) and ML to resolve this issue.

AI results are further influenced by hardware specifications and imaging protocols, in addition to variability between individual patients such as anatomic geometry, and tumor location. The impact of such variations commonly limits model accuracy and generalizability (133-135). Many AI studies lack transparency in feature selection, training, validation, and testing: even sample sizes often go unreported by many authors. To address these problems, a comprehensive guideline for the evaluation and development of reliable AI models in medical imaging (Checklist for Artificial Intelligence in Medical Imaging, CLAIM), has been proposed by Mongan et al. (129). The CLAIM framework details 42 items that can guide authors and reviewers of AI manuscripts by providing recommendations on generalizability and reproducibility for frequently encountered tasks like classification, image reconstruction, image analysis, and workflow optimization.

Improved CLAIM adherence could enhance the robustness and generalizability of trained algorithms to boost the adoption of AI in clinical practice and accelerate investigators' progress toward future innovation.

### 5.3.Future Direction

Few research papers currently focus on AI for monitoring intrafraction motion in radiotherapy compared to other AI applications, such as automatic segmentation. Moreover, current AI methods are hindered by high latency due to slow processing speed. However, with further advancements in hardware and software design, it is expected that a linear accelerator equipped for standard IGRT may be used in the near future to monitor the intrafraction motion of radiotherapy targets without additional hardware.

## 6. Conclusion

The use of AI in motion monitoring during radiotherapy has demonstrated significant success in improving accuracy compared to conventional techniques. AI methods have shown potential in markerless tracking of intrafraction movements by enhancing target visibility within onboard projection images. Although AI networks can provide accurate predictions, their decisions can be difficult to interpret, explain, debug, and validate, which poses a regulatory and ethical challenge. More research on intrafraction motion management is needed to reliably evaluate the performance of proposed AI methods.